\documentclass[sn-mathphys,Numbered]{sn-jnl}
\usepackage{graphicx}%
\usepackage{multirow}%
\usepackage{amsmath,amssymb,amsfonts}%
\usepackage{amsthm}%
\usepackage{mathrsfs}%
\usepackage[title]{appendix}%
\usepackage{xcolor}%
\usepackage{textcomp}%
\usepackage{manyfoot}%
\usepackage{booktabs}%
\usepackage{algorithm}%
\usepackage{algorithmicx}%
\usepackage{algpseudocode}%
\usepackage{listings}%
\newtheorem{Theorem}{Theorem}
\newtheorem{Definition}{Definition}
\newtheorem{Lemma}{Lemma}
\newtheorem{Conjecture}{Conjecture}

\begin{document}
	
	\title{Projectivities of informationally complete measurements}
	
	\author*[1,2]{\fnm{Hao} \sur{Shu}}\email{Hao\_B\_Shu@163.com}
	
	\affil*[1]{\orgname{Shenzhen University}}
	
	\affil[2]{\orgname{South China University of Technology}}
	
	\abstract{The physical problem behind informationally complete (IC) measurements is determining an unknown state statistically by measurement outcomes, known as state tomography. It is of central importance in quantum information processing such as channel estimating, device testing, quantum key distribution, etc. However, constructing such measurements with good properties is a long-standing problem. In this work, we investigate projective realizations of IC measurements. Conditions of informational completeness are presented with proofs first. Then the projective realizations of IC measurements, including proposing the first general construction of minimal projective IC measurements (MPICM) in no prime power dimensional systems, as well as determining an unknown state in $C^{n}$ via a single projective measurement with some kinds of optimalities in a larger system, are investigated. Finally, The results can be extended to local state tomography. Some discussions on employing several kinds of optimalities are also provided.}
	
	
	\maketitle

	\section{Introduction}

    The effectiveness of quantum channels including devices (side-channels) should be considered in the first place no matter what quantum tasks are going to be implemented. Generally speaking, channels could be estimated by comparing the input states, which are usually under control, and the corresponding output states, which might be totally unknown. Therefore, quantum state tomography, a technology for determining unknown states via the probability of measurement outcomes, is extremely important with substantial applications in quantum information processing such as channel estimating, device testing, quantum key distribution, etc. Precisely, an unknown channel can be viewed as a black box, providing unknown operators, while a series of coherent states often including a basis, are input with the corresponding output states measured for estimating purposes. Hence, the problem is determining an (usually mixed) unknown state via a large number of copies, which give the probability of measurement outcomes asymptotically. A series of measurements is called informationally complete (IC) if the probability of its outcomes determines a state uniquely. Naturally, IC measurements are in the central position of quantum state tomography.

     Research on IC measurements often have a strong connection with mutually unbiased bases (MUB)\cite{DE2010On}. It has been shown that measurements derived by total $n+1$ MUB, if they exist, are IC in $C^{n}$\cite{I1981Geometrical,WFOptimal1989}. The advantage of employing $n+1$ MUB is that they are realized by projective measurements with the minimum number of effects (In this paper, an effect always represents an operator in a POVM) and are optimal, while the disadvantage is that the existence of $n+1$ MUB is only proved for $n$ be prime powers\cite{BB2002A,D2004A} without other examples. Note that even searching 4 MUB in $C^6$ seems hopeless\cite{RL2011Mutually}. On the other hand, although non-projective measurements might not be as good as projective ones, for example in realizations, efficiency, etc, several non-projective IC measurements, such as mutually unbiased measurements (MUM)\cite{KG2014Mutually}, symmetric informationally complete (SIC) measurements\cite{RR2004Symmetric,KG2014Construction,TB2020Compounds}, minimal informationally complete (MIC) measurements\cite{DF2021The} and others\cite{TN2002Qudit}, are investigated. Besides determining general states, determining states of bounded rank\cite{V2010Information,BD2016Strictly} as well as pure ones\cite{FS2005Minimal,GC2015Five,C2021Quantum} are studied. For practical implementation, optimal choices of measurements are examined theoretically and experimentally\cite{JK2001Measurement,MB2014Optimal,IR2019Optimal,CG2021Optimal}.

     In this paper,  projective realizations of IC measurements are investigated. Firstly, in section II, we present simple but foundational sufficient and necessary conditions for informational completeness. Then in sections III and IV, we study projective realizations of IC measurements from different perspectives. Section III is dedicated to discussing minimal projective informationally complete measurements (MPICM), where we first define MPICM and demonstrate the difficulty of constructing them in $C^{n}$, but also propose a general construction for even $n$, with numerical examinations for $10\leq n\leq 100$. Although the investigation in this section is at a primary level, we illustrate its advancements and prospects in some way. In section IV, we demonstrate that states in $C^{n}$ could be determined by a single local projective measurement in $C^{n}\otimes C^{n+1}$ or a single projective measurement in $C^{n^{2}}$, both of which could be optimal in the trace sense as well as the frame potential sense. The measurements might outperform previous constructions such as SIC-POVM. In section V, we extend the arguments to state tomography with local operations in multipartite systems. Section VI provides discussions including a comparison of the general MPICM construction in section III with other bases selecting schemes in the uncertain-volume criterion as well as a doubt in employing it. Finally, there is a conclusion in section VII.

     \section{Conditions of informational completeness}

     In an n-dimensional system, a general state, or equivalently a density operator, has $n^2$ coefficients with positive semi-definite conditions and unitary trace. Not surprisingly, it can be determined by $n^2$ independent linear equations, derived by the probability of $n^2$ effects. As a sufficient condition, it has been informally employed in many discussions, such as minimal informationally complete measurements\cite{DF2021The}. However, can the number of effects be less as we only consider positive semi-definite solutions still lacks a mathematical proof. Here we formally state the following theorem with a proof presented in supplied materials.
\vspace{0.5cm}
     \begin{Theorem}
     	Let M=$\{M_{1},M_{2},...,M_{m}\}$ be a positive operator-value measurement (POVM) of $C^n$, where $M_{i}=(m^{i}_{kj})_{kj}$ is of matrix form under a computational basis, and $\bar{M}=(m^{i}_{kj})_{i,(k,j)}\in Mat_{m\times n^2}(C)$, where the $n^{2}$ columns are indexed by (k, j). Then the following statements are equivalent.
     	
     	(1) M is IC.
     	
     	(2) $rank(\bar{M})=n^2$.
     	
     	(3) There exists $N=\{N_{i}|i=1,2,...,n^2\}\subset M$ such that N is linearly independent.
     	
     	\end{Theorem}
\vspace{0.5cm}
     If we also consider non-complete POVM(i.e. do not require $\sum_{i=1}^{m}M_{i}=I$), then an IC POVM might only contain $n^{2}-1$ effects, namely $M=\{M_{i}|i=1,2,...,n^{2}-1\}$, since theoretically, the probabilities of these effects also give the probability of the effect $M_{0}=I-\sum_{i=1}^{n^{2}-1}M_{i}$, though the $n^{2}$ operators are linearly independent is still required. However, in practice, it is hard to determine how many states are lost, and thus $n^{2}-1$ effects might not be enough.

      It is easy to construct $n^2$ linearly independent rank-one effects in $C^{n}$. For example, $\{|s\rangle\langle s| \quad  |s=0,1,...,n-1\}\cup\{\frac{1}{2}(|j\rangle+|k\rangle)(\langle j|+\langle k|) \quad \ |k,j=0,1,...,n-1,k>j\}\cup\{\frac{1}{2}(|j\rangle+i|k\rangle)(\langle j|-i\langle k|) \quad |k,j=0,1,...,n-1,k>j\}$ meets the requirement.

In practical employment, the state might not be totally unknown but belongs to a known subspace $D$. For such a scenario, we define the following concept and provide an analogical theorem.
\vspace{0.5cm}
 \begin{Definition}
 	Let D be a subspace of $C^{n}$. A POVM $M=\{M_{i} |i=1,2,...,m\}$ in $C^{n}$ is said to be IC over $D$ if it is IC when measuring states in $D$.
 \end{Definition}
\vspace{0.5cm}
\begin{Theorem}
	Let D be a d-dimensional subspace of $C^n$, $B_{d}=\{|0\rangle, |1\rangle,..., |d-1\rangle\}$ be a computational basis of D and M=$\{M_{1},M_{2},...,M_{m}\}$ be a POVM of $C^n$. Then the following statements are equivalent.
	
	(1) M is IC over D.
	
	(2) There exists $N=\{N_{i}|i=1,2,...,d^2\}\subset M$ such that $N'=\{N_{i1}=\sum_{0\leq u, v\leq d-1}|u\rangle\langle u|N_{i}|v\rangle\langle v|\ |i=1,2,...,d^2\}$ is linearly independent.
\end{Theorem}
\vspace{0.5cm}
    Note that let $B_{d}=\{|0\rangle, |1\rangle,..., |d-1\rangle\}$ extend to $B_{n}=\{|0\rangle, |1\rangle,..., |d-1\rangle, |d\rangle, |d+1\rangle,..., |n-1\rangle \}$, a computational basis of $C^n$, then $N_{i1}$
	is the up-left $d\times d$ block of matrix $N_{i}$ under the computational basis. Hence, $N'=\{N_{i1}|i=1,2,...,m\}$ is exactly a POVM equivalent to POVM $N=\{N_{i}|i=1,2,...,m^2\}$ when measuring states in $D$\cite{S2020The}. In fact, $\sum_{0\leq u\leq d-1}|u\rangle\langle u|$ is the projector into $D$.

For an IC POVM $M=\{M_{i} |i=1,2,...,m\}$ over $D$, naturally, we expect that the number of effects is minimal, namely, $d^2$. The number of effects can be easily reduced to $d^2+1$ by selecting a maximal linearly independent subset in $\{M_{i1}=\sum_{0\leq u, v\leq d-1}|u\rangle\langle u|M_{i}|v\rangle\langle v|\ |i=1,2,...,m\}$, and completing via one effect. Precisely, if $\{M_{i1}=\sum_{0\leq u, v\leq d-1}|u\rangle\langle u|M_{i}|v\rangle\langle v|\ |i=1,2,...,d^{2}\}$ is such a maximal linearly independent subset, then POVM $\{M_{i} |i=1,2,...,d^{2}\}\cup\{I-\sum_{i=1}^{d^{2}}M_{i}\}$ is IC over $D$ with $d^{2}+1$ effects.

Let $M=\{M_{1},...,M_{d^2}\}$ completed by $M_{0}=I-\sum_{k=1}^{d^2}M_{k}$ be such a POVM, with $M_{01}=\sum_{k=1}^{d^2}x_{k}M_{k1}$ (viewed as an operator in $D$), where $x_{k}\in C$, $k=1,2,...,d^2$ and note that $\{M_{k1} |k=1,2,...,d^{2}\}$ (viewed as operators in $D$) is a basis of linear operators in $D$. It is obvious that there exists a $x_{k}\neq -1$, since $M_{01}+\sum_{k=1}^{d^2}x_{k}M_{k1}=I_{D}$. Without loss generality assume that $x_{1}\neq -1$, then $\{M_{0}+M_{1},M_{2},...,M_{d^2}\}$ is a IC POVM over $D$ with $d^2$ effects.

\section{Minimal projective informationally complete measurements}

Projective measurements with rank-one projectors, namely measurements derived by orthonormal bases, have their own advantages including being realized easily, highly efficient, and with low error rate. Therefore, naturally, searching projective IC measurements with minimal effects is considerable. A projective measurement $P=\{P_{1},P_{2},...,P_{s}\}$ in $C^{n}$ has at most $n$ linearly independent effects with $\sum_{i=1}^{s} P_{i}=I$ and thus for $n\geq 2$, $n$ projective measurements are not sufficient to provide $n^2$ linearly independent effects. Hence, projective IC measurements in $C^n$ contain at least $n+1$ members.
\vspace{0.5cm}
\begin{Definition}
	In $C^{n}$, a series of projective measurements is called a minimal projective informationally complete measurement (MPICM) if it contains n+1 projective measurements and is informationally complete over $C^{n}$.
\end{Definition}
\vspace{0.5cm}
Constructing projective IC measurements with as less measurements as possible in $C^{n}$ is a long-standing problem\cite{GC2015Five}, which can be partially solved by MUB. However, since the existence of $n+1$ MUB seems hopeless except when $n$ is a prime power, the problem remains open. The difficulty of constructing MPICM might be viewed via the following theorem, which will be proved in supplied materials. The main condition is similar to the existing condition of $n+1$ MUB.
\vspace{0.5cm}
\begin{Theorem}
	The following statements are equivalent in $C^n$.
	
	(1) There exists an MPICM.
	
	(2) There exists an MPICM with all projectors rank one.
	
	(3) There exists a unitary basis X of $Mat_{n\times n}(C)$ such that $X=\{I\}\cup X_{1}\cup X_{2},...,X_{n}\cup X_{n+1}$, where $X_{k}$ is a commutative set with $|X_{k}|=n-1$ for every k=1,2,...,n+1.
	
	In fact, the unitary basis in (3) can be replaced by any basis with diagonalizable operators while the number of sets in the partition indicates the number of projective measurements needed for informational completeness.
\end{Theorem}
\vspace{0.5cm}
Note that if the basis in (3) is required to be orthogonal (in matrix space), then the condition is the same as the existing condition of $n+1$ MUB\cite{BB2002A}. Therefore, constructing MPICM seems nearly as difficult as constructing $n+1$ MUB. However, examples of MPICM for $n=2r$ be even can be constructed.

 The rank-one projective measurements (bases before normalized) are constructed as follows, where the computational basis is denoted by $\{|1\rangle,|2\rangle,...,|n-1\rangle,|n\rangle\}$ while $n=2r$, $\omega=e^{\frac{2\pi i}{r}}$, and $q=e^{\frac{2\pi i}{n}}$.
\\

   \noindent r=2:
    \\
   $P_{0}=\{|1\rangle,|2\rangle,|3\rangle,|4\rangle\}$,
    \\
    \begin{small}
    $P_{1}=\{|1\rangle+|2\rangle,|3\rangle+|4\rangle, |1\rangle-|2\rangle+|3\rangle-|4\rangle,|1\rangle-|2\rangle-|3\rangle+|4\rangle\}$,
    \\
    \end{small}
    $P_{2}=\{|1\rangle+|3\rangle,|2\rangle+|4\rangle, |1\rangle-|3\rangle+i|2\rangle-i|4\rangle,$

    \qquad $|1\rangle-|3\rangle-i|2\rangle+i|4\rangle\}$,
     \\
    $P_{3}=\{|1\rangle+i|2\rangle,|3\rangle+i|4\rangle, |1\rangle-i|2\rangle+i|3\rangle+|4\rangle,$

    \qquad $|1\rangle-i|2\rangle-i|3\rangle-|4\rangle\}$,
    \\
    $P_{4}=\{|1\rangle+i|3\rangle,|2\rangle+i|4\rangle, |1\rangle-i|3\rangle+|2\rangle-i|4\rangle,$

    \qquad $|1\rangle-i|3\rangle-|2\rangle+i|4\rangle\}$.

It is not very interesting since $n+1$ MUB exist in such a case. However, a similar construction appears for $n=6$, where no $n+1$ MUB were found.

    	\begin{equation*}
    		\begin{aligned}
    r=3:
     \\
    P_{0}=\{&|1\rangle,|2\rangle,|3\rangle,|4\rangle,|5\rangle,|6\rangle\ \},
    \\
    P_{1}=\{&|1\rangle+|2\rangle,|3\rangle+|5\rangle,|4\rangle+|6\rangle,(|1\rangle-|2\rangle)+(|3\rangle-|5\rangle)+i(|4\rangle-|6\rangle),
    \\
    &(|1\rangle-|2\rangle)+\omega(|3\rangle-|5\rangle)+\omega^{2}i(|4\rangle-|6\rangle),(|1\rangle-|2\rangle)+\omega^{2}(|3\rangle-|5\rangle)+\omega i(|4\rangle-|6\rangle)\},
    \\
    P_{2}=\{&|1\rangle+|3\rangle,|2\rangle+|6\rangle,|4\rangle+|5\rangle,(|1\rangle-|3\rangle)+(|2\rangle-|6\rangle)+(|4\rangle-|5\rangle),
    \\
    &(|1\rangle-|3\rangle)+\omega(|2\rangle-|6\rangle)+\omega^{2}(|4\rangle-|5\rangle),(|1\rangle-|3\rangle)+\omega^{2}(|2\rangle-|6\rangle)+\omega (|4\rangle-|5\rangle)\},
    \\
    P_{3}=\{&|1\rangle+|4\rangle,|2\rangle+|3\rangle,|5\rangle+|6\rangle,(|1\rangle-|4\rangle)+(|2\rangle-|3\rangle)+(|5\rangle-|6\rangle),
    \\
    &(|1\rangle-|4\rangle)+\omega(|2\rangle-|3\rangle)+\omega^{2}(|5\rangle-|6\rangle),(|1\rangle-|4\rangle)+\omega^{2}(|2\rangle-|3\rangle)+\omega (|5\rangle-|6\rangle)\},
    \\
    P_{4}=\{&|1\rangle+i|2\rangle,|3\rangle+i|5\rangle,|4\rangle+i|6\rangle,(|1\rangle-i|2\rangle)+(|3\rangle-i|5\rangle)+(|4\rangle-i|6\rangle),
    \\
    &(|1\rangle-i|2\rangle)+\omega(|3\rangle-i|5\rangle)+\omega^{2}(|4\rangle-i|6\rangle),(|1\rangle-i|2\rangle)+\omega^{2}(|3\rangle-i|5\rangle)+\omega (|4\rangle-i|6\rangle)\},
    \\
    P_{5}=\{&|1\rangle+i|3\rangle,|2\rangle+i|6\rangle,|4\rangle+i|5\rangle,(|1\rangle-i|3\rangle)+(|2\rangle-i|6\rangle)+(|4\rangle-i|5\rangle),
    \\
    &(|1\rangle-i|3\rangle)+\omega(|2\rangle-i|6\rangle)+\omega^{2}(|4\rangle-i|5\rangle),(|1\rangle-i|3\rangle)+\omega^{2}(|2\rangle-i|6\rangle)+\omega (|4\rangle-i|5\rangle)\},
    \\
    P_{6}=\{&|1\rangle+i|4\rangle,|2\rangle+i|3\rangle,|5\rangle+i|6\rangle,(|1\rangle-i|4\rangle)+(|2\rangle-i|3\rangle)+(|5\rangle-i|6\rangle),
    \\
    &(|1\rangle-i|4\rangle)+\omega(|2\rangle-i|3\rangle)+\omega^{2}(|5\rangle-i|6\rangle),(|1\rangle-i|4\rangle)+\omega^{2}(|2\rangle-i|3\rangle)+\omega (|5\rangle-i|6\rangle)\}.
    \end{aligned}
   	\end{equation*}

    We would not like to construct for $n=8$ since $n+1$ MUB exist. For $n=2r\geq 10$, a general construction is presented as follows.
    \\

    For $k=1, 2..., r,\ j=1, 2,..., r$,

    $|a_{2k-1,j}\rangle=
    \left\{\begin{matrix}
    	\begin{aligned}
    	&|j\rangle+|2k+1-j\rangle    &1\leq j\leq k
    	\\
    	&|j+k\rangle+|2r+k-j\rangle  &k+1\leq j\leq r-1
    	\\
    	&|r+k\rangle+|2r\rangle      &j=r
    	\end{aligned}
    \end{matrix}\right.$

$|b_{2k-1,j}\rangle=
\left\{\begin{matrix}
	\begin{aligned}
		&|j\rangle-|2k+1-j\rangle &1\leq j\leq k
		\\
		&|j+k\rangle-|2r+k-j\rangle &k+1\leq j\leq r-1
		\\
		&|r+k\rangle-|2r\rangle &j=r
	\end{aligned}
\end{matrix}\right.$

$|c_{2k-1,j}\rangle=
\left\{\begin{matrix}
	\begin{aligned}
		&|j\rangle+q|2k+1-j\rangle &1\leq j\leq k
		\\
		&|j+k\rangle+q|2r+k-j\rangle &k+1\leq j\leq r-1
		\\
		&|r+k\rangle+q|2r\rangle &j=r
	\end{aligned}
\end{matrix}\right.$

$|d_{2k-1,j}\rangle=
\left\{\begin{matrix}
	\begin{aligned}
		&|j\rangle-q|2k+1-j\rangle &1\leq j\leq k
		\\
		&|j+k\rangle-q|2r+k-j\rangle &k+1\leq j\leq r-1
		\\
		&|r+k\rangle-q|2r\rangle &j=r
	\end{aligned}
\end{matrix}\right.$

    For $k=1, 2,..., r-1,\ j=1, 2,...r$,

    $|a_{2k,j}\rangle=
    \left\{\begin{matrix}
	\begin{aligned}
		&|j\rangle+|2k+2-j\rangle &1\leq j\leq k
		\\
		&|j+k\rangle+|2r+k-j+1\rangle &k+2\leq j\leq r
		\\
		&|k+1\rangle+|2r\rangle &j=k+1
	\end{aligned}
         \end{matrix}\right.$

$|b_{2k,j}\rangle=
\left\{\begin{matrix}
	\begin{aligned}
		&|j\rangle-|2k+2-j\rangle &1\leq j\leq k
		\\
		&|j+k\rangle-|2r+k-j+1\rangle &k+2\leq j\leq r
		\\
		&|k+1\rangle-|2r\rangle \qquad &j=k+1
	\end{aligned}
\end{matrix}\right.$

$|c_{2k,j}\rangle=
\left\{\begin{matrix}
	\begin{aligned}
		&|j\rangle+q|2k+2-j\rangle &1\leq j\leq k
		\\
		&|j+k\rangle+q|2r+k-j+1\rangle &k+2\leq j\leq r
		\\
		&|k+1\rangle+q|2r\rangle &j=k+1
	\end{aligned}
\end{matrix}\right.$

$|d_{2k,j}\rangle=
\left\{\begin{matrix}
	\begin{aligned}
		&|j\rangle-q|2k+2-j\rangle &1\leq j\leq k
		\\
		&|j+k\rangle-q|2r+k-j+1\rangle &k+2\leq j\leq r
		\\
		&|k+1\rangle-q|2r\rangle &j=k+1
	\end{aligned}
\end{matrix}\right.$
\\

$F_{i,j-r}=\sum_{k=1}^{r}\omega^{(k-1)(j-r-1)}|b_{i,k}\rangle$, $j=r+1, r+2,..., n$,

$G_{i,j-r}=\sum_{k=1}^{r}\omega^{(k-1)(j-r-1)}|d_{i,k}\rangle$, $j=r+1, r+2,..., n$.

$P_{0}=\{|j\rangle |j=1,2,...,n\}$

$P_{i}=\{|a_{i,j}\rangle, |F_{i,j}\rangle|j=1,2,...,r\}$, $i=1, 2,..., n-1$,

$P_{i}=\{|c_{i,j}\rangle, |G_{i,j}\rangle|j=1,2,...,r\}$, $i=n, n+1,..., 2n-2$.
\\

We have the following conjecture.
\vspace{0.5cm}
\begin{Conjecture}
	For $n=2r\geq 10$, measurements derived by orthogonal bases (after being normalized) $P_{0}$, $P_{1}$,..., $P_{r}$, $P_{3r-2}$, $P_{3r-1}$,..., $P_{4r-3}$ are IC, namely they form a MPICM. Moreover, we conjecture that there exists a general construction of MPICM for all $C^{n}$ (except one or two small n), in a similar way.
	\end{Conjecture}
\vspace{0.5cm}
Instead of providing a proof, we examine the conjecture with the help of Matlab, demonstrating its validity for $10\leq n=2r\leq 100$. Strangely, the construction is not suitable for $n\leq 8$. The Matlab procedures would be proposed in supplied materials. It might be the first general construction for MPICM. For odd $n$, or in fact $n=pq$, there shall be a similar construction. However, for now, we satisfy with even $n$, since a $n$-dimensional state could be viewed as a $(n+1)$-dimensional state by the natural embedding and thus could be determined by at most $n+2$ projective measurements.

The research of MPICM is in the beginning stage and we are presenting the first general construction of it. Therefore, the significance of MPICM might not be very clear. In most cases when a MUB exists, most MPICM might not be better than a MUB (but note that a MUB is also an MPICM). For example, in \cite{WFOptimal1989}, the authors stated that MUB is the optimal choice in the corresponding criterion given in that paper. The above construction might also not be necessarily better than a randomly selected $n+1$ different bases which seems to be IC in most cases (although this might lack a proof). The discussions in the criterion in \cite{WFOptimal1989} will be presented in the discussion section. However, the above works have still provided some advancements. Firstly, MPICM can be viewed as an extension of MUB but the existence and construction are more sanguine. Secondly, even if a set of randomly selected $n+1$ different orthonormal bases derives an MPICM in most cases, a general construction of MPICM for all systems is still significant, especially since an MPICM in a larger system can not be directly extended by an MPICM in a smaller system as it is not local (which will be demonstrated in section V). Thirdly, a general construction of MPICM provides a threshold to abort any worse choice under any chosen standard. Finally, the results in this section provide a framework for constructing MPICM and might inspire other significant constructions which might outperform the existing ones.

\section{Determining states via a single projective measurement}

 The goal of this section is to construct a single projective measurement in a larger system, which is IC over $C^{n}$.

\subsection{Construct by an IC POVM}

Let $M=\{M_{k}|k=1,2,...m\}$ be a IC POVM (not necessarily satisfying $\sum_{k=1}^{m}M_{k}=I$) consisting of rank-one operators, where $M_{k}=|e_{k}\rangle\langle e_{k}|$, and $|e_{k}\rangle$ are vectors in $C^n$. Let us extend the system to $n+m$ and view $\rho$ as a state in $C^{n+m}$ by the natural embedding, which has an orthogonal pure state set $\{|f_{s}\rangle|s=1,2,...,m\}$ in the orthocomplement of $C^n$.

The orthogonal basis in $C^{m+n}$ can be constructed as follow. Let

$E_{0}=\emptyset\cup\{|e_{k}^{0}\rangle=|e_{k}\rangle|k=1,...,m\}$,

$E_{1}=\{|e_{1}^{0}\rangle+|f_{1}\rangle\}\cup\{|e_{k}^{1}\rangle=|e_{k}^{0}\rangle-\langle e_{1}^{0}|e_{k}^{0}\rangle|f_{1}\rangle|k=2,3,...,m\}$,

$E_{i}=\{|e_{j}^{j-1}\rangle+|f_{j}\rangle|j=1,2,...,i\}\cup\{|e_{k}^{i}\rangle=|e_{k}^{i-1}\rangle-\langle e_{i}^{i-1}|e_{k}^{i-1}\rangle|f_{i}\rangle|k=i+1,...,m\}$.

After being normalized, $E_{m}$ consists of orthogonal states. The procedure is similar to the Schimidt orthogonalization. Hence, the (non-complete) projective measurement derived by $E_{m}$ is IC over $C^n$ while the whole system is $(n+m)$-dimensional.

To make the POVM $C^{n}$-trace balance (Here, a POVM $M=\{M_{i}|i=1,2,...,m\}$ is said to be $C^{n}$-trace balance if $\int_{\rho\in C^{n}}Tr(M_{i}\rho)=\int_{\rho\in C^{n}}Tr(M_{j}\rho)$ for all $i,j=1,2,...,m$, where $\rho$ runs over $C^{n}$ with weights and takes weighted average.), we can employ another orthogonal pure state set $\{|g_{s}\rangle|s=1,2,...,m\}$ in the orthocomplement of $C^{n+m}$ and let $E_{m}'=\{|h_{j}\rangle=|e_{j}^{j-1}\rangle+|f_{j}\rangle+x_{j}|g_{j}\rangle|j=1,2,...,m\}$, where $x_{j}\geq 0$. Hence, $x_{j}$ can be chosen such that $\frac{\int_{\rho\in C^{n}}Tr(|e_{j}\rangle\langle e_{j}|\rho)}{||h_{j}\rangle|^{2}}$, $j=1,2,...,m$, are all equal, since $\lim_{x_{j}\rightarrow \infty }||h_{j}\rangle|=\lim_{x_{j}\rightarrow \infty }||e_{j}^{j-1}\rangle+|f_{j}\rangle+x_{j}|g_{j}\rangle|=\infty$ while $\int_{\rho\in C^{n}}Tr(|e_{j}\rangle\langle e_{j}|\rho)$ is definite. If so, after being normalized, $E_{m}'$ consists of orthogonal states in $C^{n+2m}$ with derived operators $C^{n}$-trace balance. Hence, the dimension of the whole system becomes $n+2m$.

\subsection{Construct by an MPICM}

 For $M$ with minimal $n^2$ effects, the projective measurement can be local and constructed by an MPICM. Let $P_{0}=\{P_{01},P_{02},...,P_{0n}\}$, $P_{1}=\{P_{11},P_{12},...,P_{1n}\}$,..., $P_{n}=\{P_{n1},P_{n2},...,P_{nn}\}$ be a MPICM in $C^{n}$, a projective measurement in $C^{n}\otimes C^{n+1}$ could be constructed as $P=\{P_{ij}\otimes|i\rangle\langle i|\quad|i=0,1,...,n, j=1,2,...,n\}$, which is local in $C^{n}\otimes C^{n+1}$. The procedure is similar to basis selecting. $P$ is $\mathfrak{D}(C^{n})$-trace optimal as well as optimal in frame potential sense in $C^{n(n+1)}$, where $\mathfrak{D}(C^{n})$ denotes states in $C^{n}$ with the uniform distribution. The whole system is $n(n+1)$-dimensional. Hence, we have the following theorem which will be proved in supplied materials.
\vspace{0.5cm}
\begin{Theorem}
	There exists a local projective measurement consisting of rank-one projectors in $C^{n}\otimes C^{n+1}$, which is IC over $C^n$. Moreover, it can be optimal in both $\mathfrak{D}(C^{n})$-trace sense and frame potential sense in $C^{n(n+1)}$.
\end{Theorem}
\vspace{0.5cm}
 The frame potential of a set $S$ of normalized vectors in a Hilbert space is defined to be $Tr(S^2)=\sum_{|x\rangle,|y\rangle\in S}|\langle x|y\rangle|^{2}$ while $Tr(S^2)\geq max(|S|,\frac{|S|^{2}}{dim(S)})$\cite{BF2003Finite}. Let $M=\{a_{k}|e_{k}\rangle \langle e_{k}|\ |k=1,2,...,m\}$ be a POVM consisting of rank-one operators, where $S=\{|e_{k}\rangle|k=1,2,...,m\}$ is a set of pure states, then we employ $Tr(S^{2})$ as the frame potential of $M$. Frame potential is a criterion measuring the error rate of the measurement in practice. The less it is, the better the measurement is in this sense. A POVM $M=\{M_{i}| i=1,2,...,m\}$ is said to be D-trace optimal\cite{S2023Reduce}, where D is a subspace of states, if (1) $\int_{\rho\in D}Tr(M_{i}\rho)=\int_{\rho\in D}Tr(M_{j}\rho)$ for all i,j=1,2,...,m, where $\rho$ runs over D and takes average and (2) $\sum_{k=1}^{m}Tr(M_{k}\rho)=1$, for all $\rho\in D$. A trace optimal POVM should have a lower error rate caused by practical detectors.
 
Note that in $C^{n}$, the frame potential of IC POVM consisting of rank-one operators has a lower bound $n^{3}$, which is reached by SIC-POVM, while the POVM in the above construction has the frame potential $n(n+1)$ in $C^{n(n+1)}$. However, the above construction can be realized in $C^{n}$ by randomly selecting bases, namely the auxiliary partita serves as a basis selector only. Therefore, the above construction might outperform a SIC-POVM in such a criterion\footnote{The comparison of the construction and the SIC-POVM seems informal since they are in different systems. A more formal way might be naturally embedding $C^{n}$ into $C^{n(n+1)}$ and viewing the SIC-POVM of $C^{n}$ as a POVM in $C^{n(n+1)}$(One can see \cite{S2020The} for how to embed.). It becomes a non-complete POVM but with no difference when measuring states in $C^{n}$. Then comparing the frame potentials of the two POVM can be done in $C^{n(n+1)}$. Here, the frame potential of the construction in the subsection is $n(n+1)$, lower than the frame potential of SIC-POVM which is $n^{3}$.}. Also, note that a SIC-POVM can be projectively realized via an equivalent single projective POVM but the system needs to be extended to $n^{3}$ dimensions while only $n(n+1)$ dimensions are needed in the above realization (via an equivalent single projective POVM). Finally, the existence of SIC-POVM is not proven\cite{DF2021The} and the existence of MPICM, thanks to the discussions in section III, might be more expectable. Therefore, the above construction can somehow be more competitive than a SIC-POVM.

\subsection{Construct in $C^{n^{2}}$}

If we do not concern about the dimension of the auxiliary partita, then any measurement can be simply realized by a projective measurement in some extended system, known as Naimark dilation\cite{H2011Probabilistic}. However, the standard Naimark dilatation employs an auxiliary $m$-dimensional system for a POVM in $C^{n}$ with $m$ operators, leading that the whole dimension becoming $nm$, see supplied materials. Therefore, for a minimal IC POVM, the whole dimension is $n^{3}$.

Can we employ $C^{n^2}$, the minimal possible dimension, instead of $C^{nm}$? The answer is positive. In fact, any rank-one POVM $M=\{M_{k}=|e_{k}\rangle\langle e_{k}|\ |k=1,2,...,m\}$ in $C^{n}$ can be realized by a single projective measurement in $C^{m}$\cite{JK2003Entanglement}. To see this, note that the completeness condition of $M$, namely $\sum_{k=1}^{m}M_{k}=I$, is equivalent to the orthonormality of the columns of $G=\begin{pmatrix}
	|e_{1}\rangle &\\
	|e_{2}\rangle &\\
	: &\\
	: &\\
	|e_{m}\rangle
\end{pmatrix}$ (under the computational basis).
Hence, $G$ can be completed to a unitary matrix $G'$ by supplying some columns, where note that $m\geq n$ always holds for a complete rank-one POVM $M$. Finally, the rows of $G'$ derive a projective measurement in $C^{m}$ (under the computational basis extended the computational basis of $C^{n}$), which is a dilatation of $M$.

The problem left to complete the discussion is the existence of rank-one IC POVM. Such a POVM ($n^{2}$ vectors in $C^{n}$) can be constructed directly. An explicit construction which is $\mathfrak{D}(C^{n})$-trace optimal is provided in supplied materials. Some other constructions can refer to \cite{DF2021The}. Hence, we have the following theorem.
\vspace{0.5cm}
\begin{Theorem}
	For a rank-one $C^{d}$-trace optimal measurement with $d^{2}$ effects in $C^{n}$, which is IC over $C^{d}$, there exists a rank-one projective measurement in $C^{d^{2}}$, optimal in both $C^{d}$-trace sense and frame potential sense, which is IC over $C^d$. Here, applying projective tomography to states in $C^{d}$, the dimension of the whole system can be only $C^{d^{2}}$. In particular, for any positive integer $n$, there is a projective measurement in $C^{n^{2}}$, which is IC over $C^{n}$ and optimal in $\mathfrak{D}(C^{n})$-trace as well as frame potential sense.
\end{Theorem}
\vspace{0.5cm}
Note that although the standard Naimark dilatation only needs to measure the auxiliary partita whose dimension is $m$, a global unitary operator in $C^{n}\otimes C^{m}$ is required. Therefore, it might be viewed as a projective measurement in $C^{nm}$ but not in $C^{m}$.

The advantages of such a projective measurement are obvious. It obtains the minimal frame potential among all rank-one measurements that are IC over $C^{d}$ in any system, namely $d^{2}$, with the minimal auxiliary degree of freedom employed, namely operating in the minimal possible system.

The property of the dilatation depends on the property of the ordinary POVM. For example, to obtain a $C^{d}$-trace optimal rank-one projective IC measurement in $C^{d^{2}}$, there should be a rank-one $C^{d}$-trace optimal IC measurement in $C^{d}$. This can be conjectured to be true.
\vspace{0.5cm}
\begin{Conjecture}
	 For any positive integer $n$ and any weight, there exists a rank-one $C^{n}$-trace optimal measurement with exactly $n^{2}$ effects in $C^{n}$.
	\end{Conjecture}
\vspace{0.5cm}
\section{Local state tomography in multipartite systems}

In certain tasks, for example, quantum key distribution with a third party who distributes entanglement states, the quantum system is divided into separated partite. Therefore, state tomography via local operations is required. The above results can be easily generalized to local cases.

Let a multipartite system $C^{n}=\otimes_{k=1}^{K}C^{n_{k}}$, and $M^{k}$ be a IC POVM in $C^{n_{k}}$. Then $M=\otimes_{k=1}^{K}M^{k}$ is a local IC POVM in $C^{n}$ since the tensor product of linearly independent sets is linearly independent. The number of effects is $\prod_{k=1}^{K}|M^{k}|=\prod_{k=1}^{K}|n_{k}|^{2}=n^{2}$ which is minimal, if $|M_{k}|=n_{k}^{2}$ is minimal for each $k$. $M$ is $\mathfrak{D}(C^{n})$-trace optimal if each $M^{k}$ is $\mathfrak{D}(C^{n_{k}})$-trace optimal while $M$ is frame potential optimal if all $M_{k}$ are. The same argument holds for projective realizations, namely informational completeness can be realized by local projective (LP) measurements. In other words, we have the following theorem.
\vspace{0.5cm}
\begin{Theorem}
	In a multipartite system $C^{n}=\otimes_{k=1}^{K}C^{n_{k}}$, a $\mathfrak{D}(C^{n})$-trace optimal local IC measurement exists. On the other hand, an unknown state in $C^{n}=\otimes_{k=1}^{K}C^{n_{k}}$ can be determined by a LP measurement in $C^{\prod_{k=1}^{K}n_{k}^{2}}=\otimes_{k=1}^{K}C^{n_{k}^{2}}$.
\end{Theorem}
\vspace{0.5cm}
Note that a MPICM in $\otimes_{k=1}^{K}C^{n_{k}}$ can never be local. To determine an unknown state with local projective measurements, each partita $C^{n_{k}}$ must employ at least $n_{k}+1$ (rank-one, without loss generality,) projective measurements. Therefore, there are at least $\prod_{k=1}^{K}(n_{k}+1)n_{k}=n\prod_{k=1}^{K}(n_{k}+1)>n(n+1)$ effects while a MPICM has $n(n+1)$ effects.

\section{Discussion}

The article \cite{WFOptimal1989} gave a method to view a complete set of MUB as a single measurement and presented a criterion to compare different measurements by calculating the volume of the parallelepiped spanned by orthonormal bases of each subspace spanned by operators $Q_{i}=\{|x\rangle\langle x|-\frac{1}{n}I|\ |x\rangle\in P_{i}\}$, where $P_{i}$ are the selected bases\footnote{In the context in \cite{WFOptimal1989}, a larger value represents more information gain. Note that there are other terms in the criterion but only the volume depends on the measurement, and it is expected to be maximized to maximize the information. 

On the other hand, one might view a complete set of MUB (or just $n+1$ IC projective measurements) as a single measurement $\{P_{i}^{(r)}-\frac{I}{N}|i=1,2,...N,\ r=1,2,...,N+1\}$ (the notions $P_{i}^{r}$ are followed the reference article), which can be realized by projective measurements, although it is not projective. In detail, for each $r$, $\{P_{i}^{(r)}-\frac{I}{N}|i=1,2,...,N\}$ is equivalent to the measurement derived by the $r$-th basis $\{P_{i}^{(r)}|i=1,2,...,N\}$ in the context of probability, and thus the single measurement $\{P_{i}^{(r)}-\frac{I}{N}|i=1,2,...,N\}$ can be realized by the basis. Moreover, if the bases are a complete set of MUB, then $P_{i}^{(r)}-\frac{I}{N}$ is orthogonal to $P_{j}^{(k)}-\frac{I}{N}$ for $r\neq k$. Therefore, the measurement can be viewed as being realized by a projective measurement regarding a basis selection, followed by another projective measurement regarding measuring by the basis. That is also why the discussion in the third paragraph of this section, employing the criterion for a single projective measurement, is valid.}. In this section, some discussions associated with the criterion including applying to the presented constructions are provided. 

One might compare an MPICM to a MUB as well as a set of $n+1$ randomly selected bases by the method. In the case $n=4$ where a complete set of MUB exists, the volume correspondent to a complete set of MUB is 1 and to a set of $n+1$ randomly selected bases is about $10^{-6}$ while to the MPICM constructed in section III is $6.25\times10^{-2}$. In the case $n=10$ where a complete set of MUB has not been found, the volume correspondent to a set of randomly selected bases is about $10^{-30}$ while to the MPICM constructed in section III is about $2.34\times10^{-29}$. However, in the case $n=14$, the volume corresponding to a set of randomly selected bases is about $10^{-57}$ while the MPICM constructed in section III is about $1.19\times10^{-70}$. The Matlab procedures to calculate these are provided in supplied materials. Therefore, the construction in section III could perform better than a set of randomly selected bases in small systems while they might perform worse in large systems over such a criterion. On the other hand, the obviously-visible difference between the volume of MUB and the volume of randomly selected bases indicates that there might be a construction significantly improving randomly selected bases.

The comparison might also involve projective realizations via a single projective measurement in larger systems since the measurement unioning all $Q_{i}$ can be viewed as combining all bases $P_{i}$ into a single measurement as illustrated in the footnote of the last paragraph. As a single projective measurement is already a single measurement, one might directly calculate the volume of the $n^{2}-1$ operators by omitting one for a single complete projective IC measurement with $n^{2}$ operators. Hence, the volume should be 1 since all operators are orthonormal. This is not surprising since, on one hand, the construction in section IV B demonstrates that an MPICM (including a complete set of MUB) can be equivalent to a single projective measurement somehow, and on the other hand, the $n^{2}-1$ operators above can be divided into $n+1$ groups with $n-1$ operators in each such that all operators (and groups) are chosen in the same probability (providing trace optimality), thus the measurement might be viewed as a random measurement(group) selection, followed by a projective measurement, corresponding to the basis selecting and measuring in a complete set of MUB.

Similarly, the argument could be extended to other measurements such as SIC-POVM. However, for any non-orthogonal operators, the volume should be less than 1 unless they can be viewed as realizing by orthogonal ones.

\section{Conclusion}

In conclusion, we presented sufficient and necessary conditions for IC measurements and discussed the projective realizations of the IC POVM, including proposing the existence condition of MPICM, providing the first general construction of MPICM for no prime power dimensional systems, as well as realizing informational completeness over $C^{n}$ via a single projective measurement by enlarging the system. The discussions of the constructions in certain criteria are also provided and the arguments could be extended to state tomography with local measurements in a multipartite system.

  \bibliography{Bibliog}
	
\section{Supplied materials}

\subsection{Proof of theorem 1}

The equivalence of (2) and (3) is obvious since $\bar{M}$ consists of matrices in $M$ written in vector form in $C^{n^2}$. (2) implies (1) is also obvious since if the density matrix $\rho=(\rho_{ij})_{ij}$ and let $\rho'=(\rho_{11},\rho_{21},...,\rho_{n1},\rho_{12},\rho_{22},...,\rho_{n2},...,\rho_{nn})$, then $\rho'^{T}$ is the unique solution of linear equations $\bar{M}x=(Tr(M_{1}\rho),Tr(M_{2}\rho),...,Tr(M_{m}\rho))^{T}$, as $\bar{M}$ is of full rank, where $A^{T}$ denotes the transposition of $A$.

Let us prove that (1) implies (2). Surely, $rank(\bar{M})\leq n^2$. Let $\rho>0$ be a density operator, which exists obviously. If $rank(\bar{M})<n^2$, then since $\bar{M}$ is degenerated and $Tr(AB)\in R$ for Hermitian operators $A$, $B$, there exists $(y_{1},y_{2},...,y_{n^2})\neq 0$ be a solution of $\sum_{i=1}^{n^2}x_{i}Tr(M_{k}B_{i})=0, k=1,2,...m$, where $y_{i}\in R$ and $\{B_{i}|i=1,2,...,n^2\}$ is a basis of $Mat_{n\times n}(C)$, consisting of positive semi-definite operators. Hence, $\rho'=\sum_{i=1}^{n^2}y_{i}B_{i}$ is Hermitian with $Tr(\rho')=Tr(\rho'I)=Tr((\sum_{k=1}^{m}M_{k})(\sum_{i=1}^{n^2}y_{i}B_{i}))=\sum_{k=1}^{m}\sum_{i=1}^{n^2}y_{i}Tr(M_{k}B_{i})=0$. Now $\rho+s\rho'\neq\rho$ is another solution of $\bar{M}x=(Tr(M_{1}\rho),Tr(M_{2}\rho),...,Tr(M_{m}\rho))^{T}$, where $s>0$ is chosen be small enough such that $\rho+s\rho'\geq 0$. Hence, $\rho$ can not be determined by $M$ and thus, $M$ is not IC.
$\hfill\blacksquare$

\subsection{Proof of Theorem 3}

(3) implies (2): Now $Y_{k}=\{I\}\cup X_{k}$ is commutative for each $k=1, 2,..., n+1$. Therefore, there exists an orthonormal basis $B_{k}=\{|b_{kj}\rangle|j=0,1,...,n-1\}$ consisting of common eigenvectors of operators in $Y_{k}$. Let $P_{k}=\{|b_{kj}\rangle\langle b_{kj}| \quad |j=0,1,...,n-1\}$, then $P_{k}, k=1,2,...,n+1$ is a MPICM since $Span(P_{k})=Span(Y_{k})$ by the spectral decomposition of operators in $Y_{k}$.

(2) implies (1) is obvious.

(1) implies (2): Let $P_{i}=\{P_{ij}|j=0,1,...,n-1\}$, $i=0, 1,..., n$, be a MPICM, then each $P_{i}$ is commutative and thus, there exists an orthonormal basis $P_{i}'=\{|p_{ij}\rangle|j=0,1,...,n-1\}$ consisting of common eigenvectors of $P_{ij}\in P_{i}$. Let $P_{i}''=\{|p_{ij}\rangle\langle p_{ij}| \quad |j=0,1,...,n-1\}$, then $Span(\cup_{i}P_{i}'')=Span(\cup_{i}P_{i})=Mat_{n\times n}(C)$ and thus, $P_{i}''$, $i=0, 1,..., n$, consisting of rank-one operators is a MPICM.

(2) implies (3): Without loss generality, assume that $P_{n+1}=\{|p_{n+1,k}\rangle\langle p_{n+1,k}|=|k\rangle\langle k| \quad |k=0,1,...,n-1\}$, $P_{i}=\{|p_{ik}\rangle\langle p_{ik}| \quad |k=0,1,...,n-1\}$, $i=1, 2,..., n$. Let $U_{ik}=\sum_{j=0}^{n-1}\omega^{jk}|p_{ij}\rangle\langle p_{ij}|$, where $\omega=e^{\frac{2\pi i}{n}}$, $k=0, 1,..., n-1$. Then for $i=1, 2, ..., n+1$, $X_{i}=\{U_{ik}|k=1,2,...,n-1\}$ consisting of $n-1$ unitary operators is an commutative set. It remains to prove the following lemma:
\vspace{0.5cm}
\begin{Lemma}
   $X=\{I\}\cup X_{1}\cup...\cup X_{n+1}$ in above construction is linearly independent.
\end{Lemma}
\vspace{0.5cm}
\newpage
\textbf{Proof}:

$\begin{aligned}
&x_{0}I+\sum_{i=1}^{n+1}\sum_{k=1}^{n-1}x_{ik}U_{ik}
\\
=&x_{0}I+\sum_{i=1}^{n+1}\sum_{k=1}^{n-1}\sum_{s=0}^{n-1}x_{ik}\omega^{sk}|p_{is}\rangle\langle p_{is}|
=x_{0}I+\sum_{i=1}^{n+1}(\sum_{s=1}^{n-1}\sum_{k=1}^{n-1}x_{ik}\omega^{sk}|p_{is}\rangle\langle p_{is}|+\sum_{k=1}^{n-1}x_{ik}|p_{i0}\rangle\langle p_{i0}|)
\\
=&x_{0}I+\sum_{i=1}^{n+1}(\sum_{s=1}^{n-1}\sum_{k=1}^{n-1}x_{ik}\omega^{sk}|p_{is}\rangle\langle p_{is}|+\sum_{k=1}^{n-1}x_{ik}(I-\sum_{s=1}^{n-1}|p_{is}\rangle\langle p_{is}|))
\\
=&(x_{0}+\sum_{i=1}^{n+1}\sum_{k=1}^{n-1}x_{ik})I+\sum_{i=1}^{n+1}\sum_{s=1}^{n-1}\sum_{k=1}^{n-1}x_{ik}(\omega^{sk}-1)|p_{is}\rangle\langle p_{is}|
\end{aligned}$

Since $P_{i}$, $i=1, 2,..., n+1$, are IC, and $\{I\}\cup\{|p_{is}\rangle\langle p_{is}| \quad |i=1,2,...,n+1, s=1,2,...,n-1\}$ is linearly independent, $x_{0}I+\sum_{i=1}^{n+1}\sum_{k=1}^{n-1}x_{ik}U_{ik}=0$ implies $x_{0}+\sum_{i=1}^{n+1}\sum_{k=1}^{n-1}x_{ik}=0$ and $\sum_{k=1}^{n-1}x_{ik}(\omega^{sk}-1)=0$ for $i=1, 2,..., n+1$, $s=1, 2,..., n-1$. For each $i$, let $s$ runs over $1, 2,..., n-1$, then $\sum_{k=1}^{n-1}x_{ik}(\omega^{sk}-1)=0$ gives $x_{ik}=0$ for $k=1, 2,..., n-1$, since the coefficient matrix is of full rank, which can be seen as follow. The coefficient matrix $C=(\omega_{sk}-1)_{sk}$, where $s,\ k$ run over $1, 2,..., n-1$, is non-degenerated if and only if $C'=\begin{pmatrix}
	1 & 1\\
	0 & C
\end{pmatrix}$
is non-degenerated. $C'$ can be transformed into $C''=(\omega_{sk})_{sk}$, where $s$, $k$ run over $0, 1,..., n-1$, by adding the first row to others. Hence, $C'$ is non-degenerated if and only if $C''$ is non-degenerated which is true since $C''$ is a Vandermonde matrix. Finally, $x_{0}+\sum_{i=1}^{n+1}\sum_{k=1}^{n-1}x_{ik}=0$ gives $x_{0}=0$.
$\hfill\blacksquare$

\subsection{Matlab codes}

The  Matlab codes can be found at \href{https://github.com/Hao-B-Shu/IC}{https://github.com/Hao-B-Shu/IC}.

The Matlab procedures include:

(1) Calculate the rank of operators derived by the MPICM constructed in section III for $r=2$, $r=3$, and $5\leq r\leq 50$, with $n=2r$.

(2) Calculations in the discussion section.

\subsection{Proof of the orthogonality and completeness of $E_{m}$ in subsection A of section IV}

Let
$E_{0}=\emptyset\cup\{|e_{k}^{0}\rangle=|e_{k}\rangle|k=1,...,m\}$,

$E_{i}=\{|e_{j}^{j-1}\rangle+|f_{j}\rangle|j=1,2,...,i\}\cup\{|e_{k}^{i}\rangle=|e_{k}^{i-1}\rangle-\langle e_{i}^{i-1}|e_{k}^{i-1}\rangle|f_{i}\rangle|k=i+1,...,m\}$,

$F_{i}=\{|e_{j}^{j-1}\rangle+|f_{j}\rangle|j=1,2,...,i\}$, $G_{i}=\{|e_{k}^{i}\rangle=|e_{k}^{i-1}\rangle-\langle e_{i}^{i-1}|e_{k}^{i-1}\rangle|f_{i}\rangle|k=i+1,...,m\}$

Firstly, let us prove that $E_{m}$ is orthogonal. We prove by induction that for every $i$, if $|x\rangle\in F_{i}, |y\rangle\in E_{i}$ with $|x\rangle\neq |y\rangle$, then $\langle x|y\rangle=0$. The proposition is true for $i=0$. Assume that for $i-1\geq 0$, it is true. Now, we have $F_{i-1}=\{|e_{j}^{j-1}\rangle+|f_{j}\rangle|j=1,2,...,i-1\}$ and $F_{i}=F_{i-1}\cup\{|e_{i}^{i-1}\rangle+|f_{i}\rangle\}$. It is easy to show that $|e_{k}^{k-1}\rangle\in Span\{|e_{k}\rangle,|f_{1}\rangle,|f_{2}\rangle,...,|f_{k}\rangle\}$, which implies that for $j>k$, $\langle f_{j}|e_{k}^{k-1}\rangle=0$. Also $G_{i}\subset Span(G_{i-1}\cup\{|f_{i}\rangle\})$. Therefore by induction, for any $|x\rangle\in F_{i-1}$, $|y\rangle\in G_{i}$, $\langle x|y\rangle=0$. On the other hand, $\langle x|e_{i}^{i-1}\rangle=0$ since $|e_{i}^{i-1}\rangle\in G_{i-1}$. Therefore, $\langle x|(|e_{i}^{i-1}\rangle+|f_{i}\rangle)=0$. Finally, $|y\rangle\in G_{i}$ is of form $|e_{k}^{i-1}\rangle-\langle e_{i}^{i-1}|e_{k}^{i-1}\rangle |f_{i}\rangle$, which is orthogonal to $|e_{i}^{i-1}\rangle+|f_{i}\rangle$. Hence, the proposition is also true for $i$. Since $E_{m}=F_{m}$, we have that $E_{m}$ is orthogonal.

Assume that $M=\{|e_{k}\rangle\langle e_{k}| \quad |k=1,2,...,m\}$, consisting of $m$ rank-one operators, is an IC POVM (not necessarily sum to I) in $C^n$. Such a POVM always exists. Let $E_{m}=\{|e_{j}'\rangle=|e_{j}^{j-1}\rangle+|f_{j}\rangle|j=1,2,...,m\}$ as above. After being normalized, $E_{m}$ derives a (not complete) projective measurement of $C^{n+m}$. On the other hand, we can view a density operator $\rho$ in $C^n$ as a density operator in $C^{n+m}$, also denoted by $\rho$ via the natural embedding. Since $\rho\in Span\{|e_{j}\rangle\langle e_{j}| \quad |j=1,2,...,m\}$, we have $Tr(|f_{i}\rangle\langle e_{j}|\rho)=Tr(|e_{j}\rangle\langle f_{i}|\rho)=0$ and thus, $Tr(|e_{j}'\rangle\langle e_{j}'|\rho)=Tr(|e_{j}\rangle\langle e_{j}|\rho)$ determines $\rho$, $j=1,2,...,m$. Hence, The measurement is IC over $C^n$.
$\hfill\blacksquare$

\subsection{Proof of Theorem 4}

For a MPICM $P_{0}=\{P_{01},P_{02},...,P_{0n}\}$, $P_{1}=\{P_{11},P_{12},...,P_{1n}\}$,..., $P_{n}=\{P_{n1},P_{n2},...,P_{nn}\}$ in $C^{n}$, let $P=\{P_{ij}\otimes|i\rangle\langle i|\quad|i=0,1,...,n, j=1,2,...,n\}$, which is local in $C^{n}\otimes C^{n+1}$. A state $\rho$ in $C^{n}$ can be viewed as state
$\rho\otimes \frac{1}{n+1}I_{n+1}$ in $C^{n}\otimes C^{n+1}$. Now for a state $\rho$ in $C^{n}$,

$Tr((P_{ij}\otimes|i\rangle\langle i|)(\rho\otimes \frac{1}{n+1}I)) =\frac{1}{n+1}Tr(P_{ij}\rho\otimes |i\rangle\langle i|)=\frac{1}{n+1}Tr(P_{ij}\rho)$. Therefore, $P$ is IC over $C^{n}$ since $P_{i}=\{P_{i1},P_{i2},...,P_{in}\}$, $i=0,1,...,n$, is a MPICM, and thus IC, in $C^{n}$.

$\int_{\rho\in\mathfrak{D}(C^{n})}Tr((P_{ij}\otimes|i\rangle\langle i|)(\rho\otimes \frac{1}{n+1}I)) =\frac{1}{n+1}\int_{\rho\in\mathfrak{D}(C^{n})}Tr(P_{ij}\rho\otimes |i\rangle\langle i|)=\frac{1}{n+1}\int_{\rho\in\mathfrak{D}(C^{n})}Tr(P_{ij}\rho)=\frac{1}{n(n+1)}Tr(P_{ij})=\frac{1}{n(n+1)}$ are equal for all $i, j$. Therefore, $P$ is $\mathfrak{D}(C^{n})$-trace balance. And since $\sum_{i=0}^{n}\sum_{j=1}^{n}\int_{\rho\in\mathfrak{D}(C^{n})}Tr((P_{ij}\otimes|i\rangle\langle i|)(\rho\otimes \frac{1}{n+1}I)) =\frac{1}{n+1}\sum_{i=0}^{n}\int_{\rho\in\mathfrak{D}(C^{n})}Tr(\sum_{j=1}^{n}P_{ij}\rho)=1$, $P$ is $\mathfrak{D}(C^{n})$-trace optimal.

The frame potential of $P$ is $|P|=n+n^{2}$ which is minimal in $C^{n}\otimes C^{n+1}$, since $P$ is a complete rank-one projective measurement with $n+n^{2}$ effects.
$\hfill\blacksquare$

\subsection{The standard Naimark dilatation}

For a POVM $M=\{M_{k} |k=0,1,...,m-1\}$ in $C^{n}$, the standard Naimark dilatation is as follow. There is a unitary operator $U$
in $C^{n}\otimes C^{m}$, where $C^{n}$ is the ordinary system and $C^{m}$ is the auxiliary system, such that $U(|\phi\rangle|0\rangle)=\sum_{k=0}^{m-1}M_{k}|\phi\rangle|k\rangle$. Hence, measuring a state in $C^{n}$ via $M$ is equivalent to measuring the state operated by $U$ after being assisted with an auxiliary state $|0\rangle$ via projective measurement $\{|k\rangle\langle k|\ |k=0,1,...,m-1\}$ in the auxiliary system $C^{m}$.

\subsection{A construction of rank-one projective IC measurement in $C^{n}$}

Firstly, since $C^{n_{1}n_{2}}\cong C^{n_{1}}\otimes C^{n_{2}}$ while the tensor product of rank-one IC POVM in $C^{n_{1}}$ and $C^{n_{2}}$ is a rank-one IC POVM in $C^{n_{1}}\otimes C^{n_{2}}$, we only need to construct for prime $n$.

Therefore, the goal is constructing vectors $|e_{k}\rangle\in C^{n}, k=1,2,...,n^{2}$ such that the columns of $G=\begin{pmatrix}
	|e_{1}\rangle &\\
	|e_{2}\rangle &\\
	: &\\
	: &\\
	|e_{n^{2}}\rangle
\end{pmatrix}_{n^{2}\times n}$
 are orthonormal with linearly independent $|e_{k}\rangle\langle e_{k}|, k=1,2,...,n^{2}$.

 \begin{Lemma}
 	Let $G=\begin{pmatrix}
 		|e_{1}\rangle &\\
 		|e_{2}\rangle &\\
 		: &\\
 		: &\\
 		|e_{m}\rangle
 	\end{pmatrix}_{m\times n}$, and
  $G\times diag(b_{1},b_{2},...,b_{n})_{n\times n}=\begin{pmatrix}
 	|e_{1}'\rangle &\\
 	|e_{2}'\rangle &\\
 	: &\\
 	: &\\
 	|e_{m}'\rangle
 \end{pmatrix}_{m\times n}$, where $b_{j}\in C-\{0\}, j=1,2,...,n$.
If $|e_{k}\rangle\langle e_{k}|, k=1,2,...,m$ are linearly independent, then $|e_{k}'\rangle\langle e_{k}'|, k=1,2,...,m$ are linearly independent.
 \end{Lemma}
\vspace{0.5cm}
 \textbf{Proof:}

 Let $|e_{k}\rangle=\sum_{j=0}^{n-1}x_{kj}|j\rangle$, then  $|e_{k}'\rangle=\sum_{j=0}^{n-1}x_{kj}b_{j}|j\rangle$.
 \\
   Hence, $\sum_{k=1}^{m}a_{k}|e_{k}'\rangle\langle e_{k}'|=\sum_{k=1}^{m}a_{k}\sum_{j,j'=0}^{n-1}x_{kj}\overline{x_{kj'}}b_{j}\overline{b_{j'}}|j\rangle\langle j'|=\sum_{j,j'=0}^{n-1}(\sum_{k=1}^{m}a_{k}x_{kj}\overline{x_{kj'}}b_{j}\overline{b_{j'}})|j\rangle\langle j'|$. Therefore,

$\sum_{k=1}^{m}a_{k}|e_{k}'\rangle\langle e_{k}'|=0\Leftrightarrow \sum_{k=1}^{m}a_{k}x_{kj}\overline{x_{kj'}}b_{j}\overline{b_{j'}}=0$ for all $j, j'$ $\Leftrightarrow \sum_{k=1}^{m}a_{k}x_{kj}\overline{x_{kj'}}=0$ since $b_{j}\neq0$, for all $j, j'$
 \\
  $\Leftrightarrow \sum_{j,j'=0}^{n-1}\sum_{k=1}^{m}a_{k}x_{kj}\overline{x_{kj'}}|j\rangle\langle j'|=\sum_{k=1}^{m}\sum_{j,j'=0}^{n-1}a_{k}x_{kj}\overline{x_{kj'}}|j\rangle\langle j'|=\sum_{k=1}^{m}a_{k}|e_{k}\rangle\langle e_{k}|=0 \Leftrightarrow a_{k}=0$ for all $k$ since $|e_{k}\rangle\langle e_{k}|, k=1,2,...,m$ are linearly independent. This proves that $|e_{k}'\rangle\langle e_{k}'|, k=1,2,...,m$ are linearly independent.
 $\hfill\blacksquare$
\\

Thanks to Lemma 2, we only need to construct vectors $|e_{k}\rangle\in C^{n}$ such that the columns of $G$ are nonzero and orthogonal with linearly independent $|e_{k}\rangle\langle e_{k}|, k=1,2,...,n^{2}$, followed by normalizing the columns.

For $n=2$, $|e_{1}\rangle=|0\rangle+|1\rangle$, $|e_{2}\rangle=|0\rangle-|1\rangle$, $|e_{3}\rangle=|0\rangle+x|1\rangle$, $|e_{4}\rangle=|0\rangle-x|1\rangle$, where $x\notin R$ and $|x|\neq 1$ satisfy the requirements.

If $n\geq 3$ is odd, let $|e_{lj}\rangle=\sum_{k=0}^{n-1}a_{lk}w^{jk}|k\rangle, l, j=0,1,...,n-1, w=e^{\frac{2\pi i}{n}}$, then the columns of $|e_{lj}\rangle$ are orthogonal. For simplicity, assume that $a_{lk}=\delta_{lk}(x_{l}-1)+1=
\left\{\begin{matrix}\begin{aligned}
	  1   \quad & l\neq k\\
      x   \quad & l=k	
    \end{aligned}
\end{matrix}\right.$, where $x\in R$, $x\neq 1, -1, 1-\frac{n}{2}$. Then $|e_{lj}\rangle\langle e_{lj}|$ are linearly independent. Hence, $|e_{lj}\rangle, l, j=0,1,...,n-1$ satisfy the requirements since the columns of $G$ are obviously nonzero.

We should prove the following lemma.
\vspace{0.5cm}
\begin{Lemma}
	The $|e_{lj}\rangle\langle e_{lj}|,l,j=0,1,...,n-1$ constructed above are linearly independent.
\end{Lemma}
\vspace{0.5cm}
\newpage
 \textbf{Proof:}
 \\

 $|e_{lj}\rangle\langle e_{lj}|=\sum_{k,k'=0}^{n-1}a_{lk}\overline{a_{lk'}}w^{j(k-k')}|k\rangle\langle k'|$. Therefore,
 \\
 $|e_{lj}\rangle\langle e_{lj}|$ are linearly independent $\Leftrightarrow$ the matrix $A=(a_{lk}\overline{a_{lk'}}w^{j(k-k')})_{(l,j),(k,k')}\in Mat_{n^{2}\times n^{2}}(C)$ is non-degenerated.

 Let us write $A=(a_{lk}\overline{a_{lk'}}w^{j(k-k')})_{(l,j),(k,k')}=(A_{lk})_{l,k}$ in block form, where $A_{lk}=(a_{lk}\overline{a_{lk'}}w^{j(k-k')})_{j,k'}\in Mat_{n\times n}(C)$. By reordering the columns of $A$, namely reordering the columns of the blocks $A_{lk}$ by putting its $k$-th column as its first column, followed by the $(k+1)$-th, $(k+2)$-th, ... columns and ended by the $(k+n-1)$-th column, where the sums are in the $mod(n)$ sense, the matrix becomes $(A_{lk}')_{l,k}$, where
 \\
 $A_{lk}'=\begin{pmatrix}
 	a_{lk}\overline{a_{lk}}(w^{0})^{0} &a_{lk}\overline{a_{l\ k+1}}(w^{0})^{-1} & ... &a_{lk}\overline{a_{l\ k+n-1}}(w^{0})^{-(n-1)} \\
 	a_{lk}\overline{a_{lk}}(w^{1})^{0} &a_{lk}\overline{a_{l\ k+1}}(w^{1})^{-1} & ... &a_{lk}\overline{a_{l\ k+n-1}}(w^{1})^{-(n-1)} \\
 	 &... & ... & \\
 	 a_{lk}\overline{a_{lk}}(w^{n-1})^{0} &a_{lk}\overline{a_{l\ k+1}}(w^{n-1})^{-1} & ... &a_{lk}\overline{a_{l\ k+n-1}}(w^{n-1})^{-(n-1)}
  \end{pmatrix}$
 	 \\
 	 $=V\times diag(a_{lk}\overline{a_{lk}},a_{lk}\overline{a_{l\ k+1}},...,a_{lk}\overline{a_{l\ k+n-1}})$, and $V=\begin{pmatrix}
 	 	(w^{0})^{0} &(w^{0})^{-1} & ... &(w^{0})^{-(n-1)} \\
 	 	(w^{1})^{0} &(w^{1})^{-1} & ... &(w^{1})^{-(n-1)} \\
 	 	&... & ... & \\
 	 	(w^{n-1})^{0} &(w^{n-1})^{-1} & ... &(w^{n-1})^{-(n-1)}
 	 \end{pmatrix}$ is non-degenerated.

Hence,
\\
$A$ is non-degenerated $\Leftrightarrow$
\\
$(A_{lk}')_{lk}=\begin{pmatrix}
	V & 0   & 0   & ... & 0 \\
	0 & V   & 0   & ... & 0 \\
	  & ... & ... & ... &   \\
	0 & 0   & 0   & ... & V
\end{pmatrix}W$ is non-degenerated, where $W=(W_{lk})_{lk}$, $W_{lk}=diag(a_{lk}\overline{a_{lk}},a_{lk}\overline{a_{l\ k+1}},...,a_{lk}\overline{a_{l\ k+n-1}})$ $\Leftrightarrow$ $W$ is non-degenerated.

After reordering the columns and rows, $W$ can be transformed into $W'=\begin{pmatrix}
	T_{0} & 0 & 0 &... & 0\\
	0 & T_{1} & 0 &... & 0\\
	  & ... & ... &... & \\
	0 & 0 & 0 &... & T_{n-1}
\end{pmatrix}$, where $T_{t}=(a_{lk}\overline{a_{l\ k+t}})_{lk}\in Mat_{n\times n}(C)$. Hence, $A$ is non-degenerated $\Leftrightarrow$ $T_{t}, t=0,1,...,n-1$ are all non-generated.

 By taking $a_{lk}=\delta_{lk}(x_{l}-1)+1=
\left\{\begin{matrix}\begin{aligned}
		1   \quad & l\neq k\\
		x   \quad & l=k	
	\end{aligned}
\end{matrix}\right.$, we have $T_{t}=T-(1-\overline{x})(X^{\dagger})^{n-t}$ for $t\neq 0$, where $T=\sum_{k=1}^{n-1}X^{k}+xI$, $X$ is the Pauli matrix, as usual. Hence, $T$ commutes with $(1-\overline{x})(X^{\dagger})^{n-t}$. Therefore, the eigenvalues of $T_{t}$ are the minus of the corresponding eigenvalues of $T$ and $(1-\overline{x})(X^{\dagger})^{n-t}$. To make $T_{t}$ non-degenerated, we only need to take $x$ such that $T$ and $(1-\overline{x})(X^{\dagger})^{n-t}$ have no common eigenvalues. An easy calculation shows that the eigenvalues of $T$ are $x-1$ and $x-n+1$, while the eigenvalues of $(1-\overline{x})(X^{\dagger})^{n-t}$ are contained in $\{(1-\overline{x})w^{k} |k=0,1,...,n-1\}$, just note that the eigenvalues of $X$ are $n$-th roots of unit.

By letting $x\in R$ and $x\neq 1, 1-\frac{n}{2}$, $T_{t}, t=1,2,...,n-1,$ are all non-degenerated. On the other hand, the eigenvalues of $T_{0}$ are $|x|^{2}-1$ and $|x|^{2}+n-1$. By letting $|x|\neq 1$, $T_{0}$ is non-degenerated. Therefore, for $x\in R-\{1,-1,1-\frac{n}{2}\}$, $A$ is non-degenerated, which implies that $|e_{lj}\rangle\langle e_{lj}|$ constructed above are linearly independent.
$\hfill\blacksquare$

On the other hand, to normalize the columns, all columns are multiplied by the same coefficient $\frac{1}{\sqrt{n|x|^{2}+n^{2}-n}}$. Hence, finally, $|e_{lj}\rangle=\frac{1}{\sqrt{n|x|^{2}+n^{2}-n}}\sum_{k=0}^{n-1}(\delta_{lk}(x-1)+1)w^{jk}|k\rangle$. Therefore $Tr(|e_{lj}\rangle\langle e_{lj}|)=\frac{1}{n}$ are all equal.
$\hfill\blacksquare$

	\end{document}